\documentclass{PoS}

\title{Annual Variations of the Galactic Cosmic Ray Intensity and Seasonal Distribution of the Cloudless Days and Cloudless Nights in Abastumani (41.75$^{o}$N, 42.82$^{o}$E; Georgia): (1)  experimental study and
 (2) theoretical modeling
}

\ShortTitle{Annual Variations of Cosmic Rays}

\author{\speaker{M. V. Alania }\thanks{A footnote may follow.}\\
        Inst.  Math. and Physics, Siedlce University, Siedlce, Poland\\ Institute of Geophysics, Tbilisi State University, Tbilisi, Georgia
\\
        E-mail: \email{alania@uph.edu.pl}}

\author{G. G. Didebulidze \\
        E. Kharadze Abastumani Astroph. Observatory, Ilia State University, Tbilisi, Georgia\\
       E-mail: \email{didebulidze@iliauni.edu.ge}}
\author{R. Modzelewska  \\
        Inst.  Math. and Physics, Siedlce University, Siedlce, Poland\\
       E-mail: \email{renatam@uph.edu.pl}}
\author{M. Todua   \\
        E. Kharadze Abastumani Astroph. Observatory, Ilia State University, Tbilisi, Georgia\\
       E-mail: \email{mayatodua@iliauni.edu.ge}}       
\author{A. Wawrzynczak  \\
        Inst.  of Computer Science, Siedlce University, Siedlce, Poland\\
       E-mail: \email{awawrzynczak@uph.edu.pl}}       
       
\abstract{We study a possible relationships between seasonal distributions of the visually observed cloudless days (CD) and cloudless nights (CN) at Abastumani Astrophysical Observatory (41.75N, 42.82E; Georgia) in 1957-1993. The annual variations of monthly numbers of CD and CN have been observed, with maximum in August for CD and in September for CN. During geomagnetic disturbances it is also observed the growth of number of CD in September and March (equinoctial months), and for CN, together with September, in June, April and February. We assume that this phenomenon indicates an influence of cosmic factors on cloudiness, as well as the existence of semiannual and possibly shorter-periodicity variations. This cosmic factor can be the manifestation of different rates of the galactic cosmic rays (GCRs) flux variations in CD and CN periods. The influence of GCR flux on ionization of lower atmosphere and variations of density of cloud condensation nuclei also can be connected to the annual and seasonal changes of temperature at Earth surface of this region. To comprehend behaviors of the annual and semi-annual variations of the GCR intensity and their possible relationships with the seasonal distributions of CD and CN we compose and numerically solve two dimensional (2-D) time dependent transport equation including all important processes in the heliosphere. An analysis of experimentally observed and theoretically obtained results have been carried out.}

\FullConference{The 34th International Cosmic Ray Conference,\\
		30 July- 6 August, 2015\\
		The Hague, The Netherlands}

\begin{document}

\section{INTRODUCTION}
A relationship of changes of the cloudless days (CD) and cloudless nights (CN) numbers  with the GCR intensity variations is studied at first time based on the unique observational data in the Abastumani Astrophysical Observatory (Georgia) during 1957-1993. So, in the beginning we consider the behavior of the average GCR intensity during the yearly period based on neutron monitors data. Then we compare results with similar changes in CD and CN for the same period. The inter-annual changes of atmospheric processes, including cloud covering in the lower atmosphere, can possibly be related to the seasonal changes of absorption of solar electromagnetic radiation energy by the Earth's surface, which depend on geometry of solar-terrestrial position in the heliosphere. During Earth's rotation around the Sun, the interplanetary and geomagnetic field geometry and, as a result, the influence of the solar corpuscular radiation on the events on the magnetosphere \cite{russell73} and also the magnetosphere-ionosphere-atmosphere coupling change \cite{cnossen12}. To generate geomagnetic disturbances by solar wind, the interplanetary magnetic field (IMF) and geomagnetic field configuration are effective at equinox months (March/April and September/October) \cite{russell73}, where the $z$ component of the IMF is directed southward. The high number of geomagnetic disturbances at equinoxes and a small number around solstice months gives semi-annual variability of frequency of appearance of the phenomena characteristic for magnetosphere-ionosphere-atmosphere coupling processes \cite{TD13}. Solar wind disturbances cause changes not only in the geomagnetic field but modulate GCR flux, as well. These in turn can cause variations of ions produced by GCR and hence the density of cloud condensation nuclei (CCN). Thus, there is a possible coupling between changes in GCR flux and cloud formation processes \cite{Svensmark97}-\cite{Marsh00}.
High mean day-night temperature in the lower atmosphere and the Earth's surface can be favorable for cloudless days and nights. When a difference between day and night-time temperature is comparatively large, then cloud formation conditions are favorable, including the influence of cosmic factors, like GCR. Seasonal peculiarities of temperature variations  from day to the night as well as with height for given region of the lower atmosphere also can change the annual and semi-annual effect of production of CCN by GCR. This in turn may result in different behavior of cloud formation during the day and the night.

Increase of the CCN density produced by GCRs flux enhancement, where the water vapor is near saturation, should stimulate the cloud formation processes and grow the cloudiness \cite{Tinsley06}. Decrease of GCRs flux in similar conditions should be favorable for cloudless days and nights. In the lower atmosphere where the clouds mostly are formed, the temperature and humidity variations are different for day and night. Because of this, the processes initiating cloud formation by GCR may be different during day and the night. Together with seasonal as well as day-night variations of atmospheric conditions, it is possible that the impact of cosmic factors on cloud covering change by day-night and seasons.
Together with the long-term variations of cloud cover, which can correlate with solar and geomagnetic activities, as well as galactic cosmic ray flux \cite{Svensmark97}-\cite{Marsh00},  it should be important to study their inter-annual/seasonal variations, which is the main purpose of this paper. We will examine inter-annual distribution of CD and CN for various level of geomagnetic activity and variations in it caused by possible cosmic factor. We also consider the inter-annual variations of the main cosmic factors: GCRs flux, solar wind velocity, and IMF at the solar activity minimum (1975-1978) and maximum (1990-1991) phases.  The inter-annual distribution of GCRs flux will be considered theoretically as well. Numerically will be solved two-dimensional (2-D) time-dependent transport equation including all relevant processes in the heliosphere. An analysis of experimentally observed and theoretically obtained results have been carried out.

\begin{figure}[tbp]
  \begin{center}
\includegraphics[width=0.55\hsize]{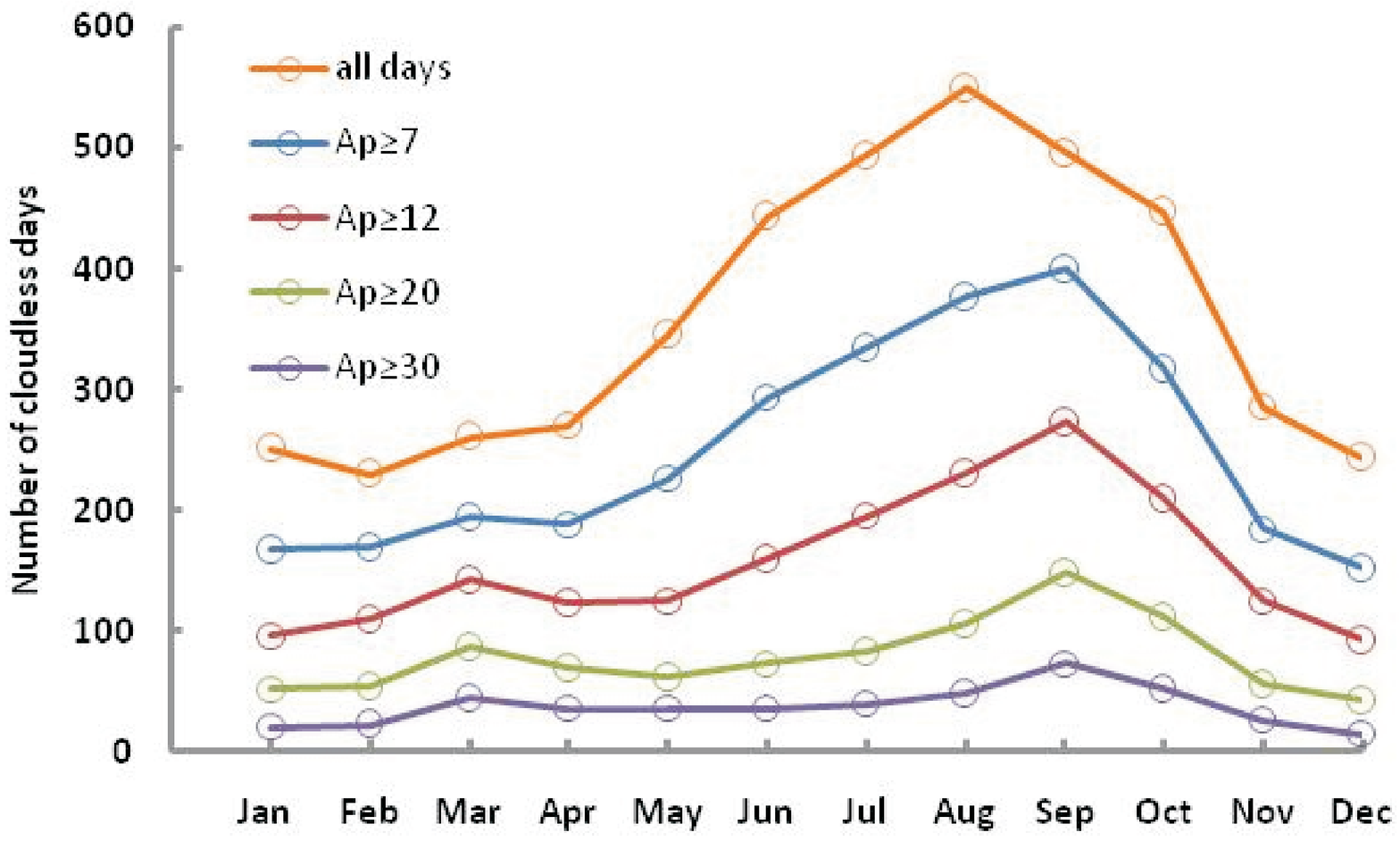}
\end{center}
\caption{\label{fig:1} Inter-annual distribution of total monthly numbers of cloudless days in Abastumani Astrophysical Observatory in 1957-1993 at different geomagnetic disturbances: for $Ap\geq7$ (blue circles), $Ap\geq12$ (red), $Ap\geq20$ (green) and $Ap\geq30$ (violet). Yellow line corresponds to all cloudless days. }
\end{figure}

\begin{figure}[tbp]
  \begin{center}
\includegraphics[width=0.55\hsize]{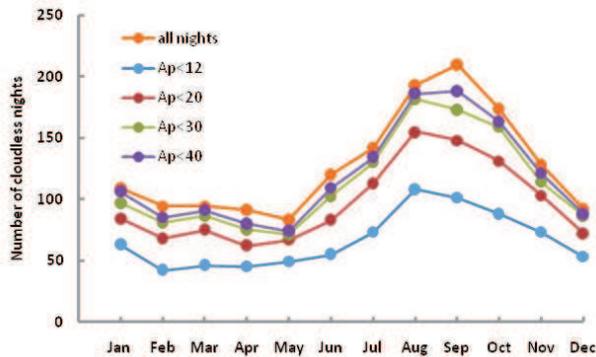}
\end{center}
\caption{\label{fig:2} Inter-annual distribution of total monthly numbers of cloudless nights in Abastumani Astrophysical Observatory in 1957-1993 at different geomagnetic disturbances: for $Ap<12$ (blue dot), $Ap<20$ (red), $Ap<30$ (green) and $Ap<40$ (violet). Yellow line corresponds to all cloudless nights.}
\end{figure}

\section{Experimental data}
Figure~\ref{fig:1} shows that the biggest number of cloudless days is in August. During magnetically relatively disturbed conditions ($Ap\geq7$) it shifts to September, where is observed a maximal number of magnetically disturbed day-nights \cite{russell73}. In the Fig.~\ref{fig:1}  the increase of number of magnetically disturbed cloudless days is observed, both in September and March, pointing to semi-annual variations modulating the annual variations of cloudless days, which indicates the influence of cosmic factors on cloud covering. This phenomenon is enforced by the fact that maximal number of cloudless nights is in September (Fig.~\ref{fig:2}) and at geomagnetically comparatively quiet periods ($Ap<40$) it is in August again, like for cloudless days. We note that in this region the maximal day-night temperature at the surface is observed mostly in August and, as was mentioned, the number of cloudless day-nights is more expectable. During magnetically disturbed conditions, such annual distribution is modulated by semi-annual variations that can be caused by the influence of cosmic factors on cloud covering. We assume that one of these cosmic factors, influencing cloud covering processes, can be considered the inter-annual variations of GCRs flux at cloudless days and nights.
\section{Model of the annual GCR intensity variation}
We assume that as one of the conventional candidates to affect the character of cloud formation and cloud coverage (cloudiness and cloudless processes) can be considered GCRs. An energy contribution of  cosmic rays in the Earth atmosphere is smaller than the energy from a total solar radiation. Nevertheless, cosmic ray particles are only a source of ionization in the lower atmosphere \cite{Tinsley06}, being further a reason for cloud formation. Unfortunately, there is not yet  well-known  how the  ionization produced by cosmic ray particles takes place in microphysics of the cloud formation.
\begin{figure}[tbp]
  \begin{center}
\includegraphics[width=0.55\hsize]{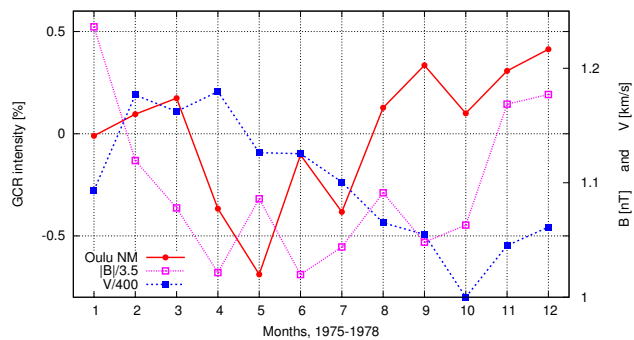}
\end{center}
\caption{\label{fig:3} Superimposed over years 1975-1978 monthly changes of the solar wind velocity $V$ and the interplanetary magnetic field strength $B$ and corresponding variations in the GCR intensity based on the Oulu neutron monitor data. To fit one scale, the solar wind velocity was scaled down by 400 km/s and IMF strength by 3.5 nT. }
\end{figure}

\begin{figure}[tbp]
  \begin{center}
\includegraphics[width=0.55\hsize]{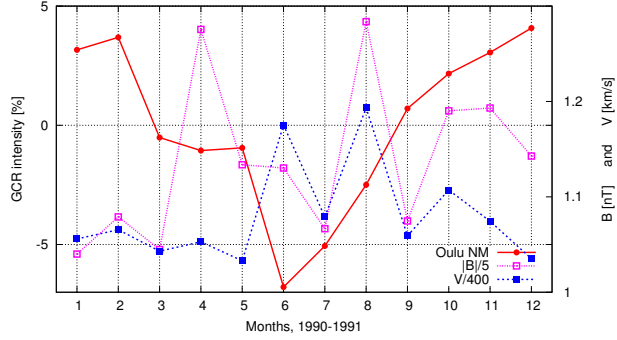}
\end{center}
\caption{\label{fig:4} Superimposed over years 1990-1991 monthly changes of the solar wind velocity $V$ and the interplanetary magnetic field strength $B$ and corresponding variations in the GCR intensity based on the Oulu neutron monitor data. To fit one scale the solar wind velocity was scaled down by 400 km/s and IMF strength by 5 nT.}
\end{figure}

Distributions of CD and CN have clearly expressed annual variation (Fig.~\ref{fig:1} and Fig.~\ref{fig:2}). If we ascribe these variations to the cloud formation owing to the ionization by cosmic ray particles (of energy 5-35 GeV, to which neutron monitor (NM) respond), there is possible to estimate an expected annual changes of cosmic ray flux solving Parker's transport equation of anisotropic diffusion \cite{Parker65}. We believe that any model describing annual changes of the GCR intensity giving an opportunity to estimate a coupling between cosmic ray flux and the cloud formation is extremely useful. In  \cite{Alania13} we have solved Parker's transport equation installing in equation the annual and semi-annual alternation of solar wind velocity comparable with experimental data. We demonstrated  that changes in solar wind velocity  in the range $\sim$ 350-500km/s cause about 0.3\% changes in the GCR intensity. This variation could not provide the enough changes of the level of ionization (aerosols) responsible for cloudiness-cloudless state of the Earth's atmosphere \cite{Alania13} and references therein.

In this paper, we compose a new 2-D time-dependent model of GCR propagation in the heliosphere including two crucial parameters i.e. in-situ measurements of the solar wind velocity $V$ and strength $B$ of the IMF. Besides, in the present paper we consider one minimum and one maximum epoch of solar activity, separately. For this purpose we performed the superposition of the monthly changes of the solar wind velocity $V$ and the IMF strength $B$ during three years in the minimum 1975-1978 of solar activity and two years 1990-1991 in the maximum of solar activity. Fig.\ref{fig:3} presents superimposed over years 1975-1978 monthly changes of the solar wind velocity $V$ and the IMF strength $B$ and corresponding variations in the GCR intensity based on the Oulu NM data. The same for the period of maximum of solar activity 1990-1991 is presented in Fig.~\ref{fig:4}. The selection of Oulu NM with the effective rigidity of 10-12 GV is caused by: (1) Oulu NM operates  stable  for the extended period, (2) data of Oulu NM does not undergone changes due to disturbances of the Earth magnetosphere, (3) changes in   amplitudes of various  classes of the GCR intensity variations are relatively significant, and (4) data of Oulu NM  are sensitive to  any kind of global changes in heliosphere.

Fig.~\ref{fig:3} shows that in the minimum epoch 1975-1978 an average annual alternations (1) of solar wind velocity is $\sim$ 100 km/s ( from $\sim$ 400 up to $\sim$ 500 km/s), (2) of the IMF strength B is $\sim$ 1.2 nT ( from $\sim$ 5.05 up to $\sim$ 6.25nT), and (3) in cosmic ray variations a range of changes is,  I $\sim$ 1.25\%. We underline that there hardly can be recognized  any definite  relationship (inter-annual or annual) among the changes of the parameters - $V$, $B$ and $I$.
Fig.~\ref{fig:4} shows that in maximum  epoch 1990-1991 an average annual alterations (1) of solar wind velocity is $\sim$ 90 km/s (from $\sim$ 410 up to $\sim$480 km/s), (2) of the IMF strength B is $\sim$1.0 nT ( from $\sim$ 5.1 up to $\sim$6.1nT), and (3) in cosmic ray variations a range of changes is,  $I$ $\sim$ 11\%. We underline that there hardly can be recognized  any precise regular (inter-annual or annual) relationships among the changes of these parameters - $V$, $B$ and $I$.
Presented in the Fig.~\ref{fig:3} and Fig.~\ref{fig:4} monthly changes of the solar wind velocity $V$ and IMF strength $B$ were implemented into the Parker's time-dependent 2-D transport equation \cite{Parker65}:
\begin{eqnarray}
\label{ParkerEq}
\frac{\partial N}{\partial\tau}=\nabla\cdot (K_{ij} ^{S}\cdot \nabla N)-(v_{d}+U)\cdot \nabla N+\frac{R}{3}\frac{\partial N}{\partial R}\nabla U,
\end{eqnarray}
where, $N$ and $R$ are omnidirectional distribution function and rigidity of GCR particles, respectively;  $\tau$-time, $V$-solar wind velocity,  $v_{d}$ is the drift velocity. We set up the dimensionless density $f=\frac{N}{N_{0}}$ , $t=\frac{\tau}{\tau_{0}}$ time   and  $r=\frac{\rho}{\rho_{0}}$ distance; where, $N_{0}$ is density in the Local Interstellar Medium (LISM) accepted as $N_{0}=4\pi I_{0}$, the intensity $I_{0}$ in the LISM \cite{Webber} has the form:   $I_{0}=21.1T^{-2.8}/(1+5.85T^{-1.22}+1.18T^{-2.54})$; $T$ is kinetic energy in GeV; $\rho$  and $\rho_{0}$   are the radial distance and size of the modulation region;  $\tau_{0}$  -is the characteristic time corresponding to the changes in heliosphere for the particular class of GCR variation. For the annual variation, we accept that $\tau_{0}$  is equal to one year. A size of the modulation region is $\rho_{0}$=100AU, the upwind-downwind asymmetry of the heliosphere is not taken into account, as far $\rho_{0}$  is significantly greater (more than 10 times) than the Larmor radius of GCR particles to which NMs respond.

\begin{figure}[tbp]
  \begin{center}
\includegraphics[width=0.55\hsize]{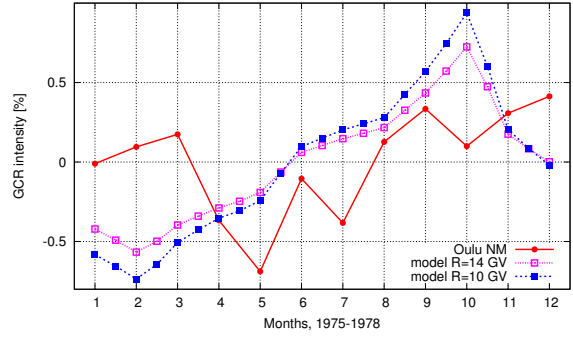}
\end{center}
\caption{\label{fig:5} The GCR intensity variation for the Oulu neutron monitor (red line)
and expected from modeling for rigidity $R= 14$ and $10$ GV (magenta and blue line) results for the period of solar activity minimum.}
\end{figure}

\begin{figure}[tbp]
  \begin{center}
\includegraphics[width=0.55\hsize]{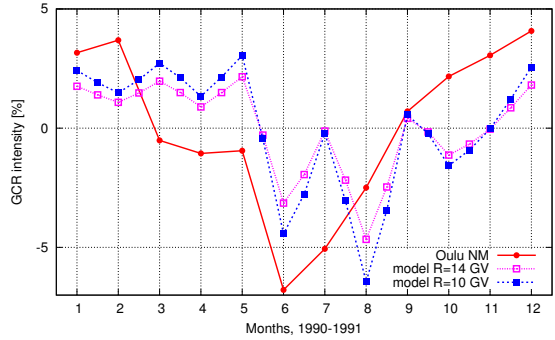}
\end{center}
\caption{\label{fig:6} The GCR intensity variation for the Oulu NM (red line) and results
 expected from modeling for rigidity R= 14 and 10 GV (magenta and blue line) for the period of solar activity maximum. }
\end{figure}

The anisotropic diffusion tensor of GCR $K_{ij}=K_{ij} ^{(S)}+K_{ij} ^{(A)}$ consists of the symmetric
$K_{ij} ^{(S)}$   and $K_{ij} ^{(A)}$  -antisymmetric parts. We implement a drift velocity of GCR particles as, $v_{d,i}=\frac{\partial K_{ij} ^{(A)}}{\partial x_{j}}$ \cite{jokipii77}.
The heliospheric magnetic field vector  $B$ is taken, as  $B=(1-2H(\theta-\theta^{'}))(B_{r}\cdot e_{r}+B_{\varphi} e_{\varphi})$ \cite{kopriva79}-\cite{kota83},
   where $H$ is the Heaviside step function changing the sign of the global magnetic field in each hemisphere and $\theta^{'}$  corresponds to the heliolatitudinal position of the heliospheric neutral sheet (HNS), $e_{r}$  and $e_{\varphi}$  are the unite vectors directed along the components $B_{r}$  and  $B_{\varphi}$ of the IMF for the two-dimensional Parker field \cite{Parker58}. Parker's spiral heliospheric magnetic field is implemented through the angle  $\psi=arctan(-B_{\varphi}/B_{r})$ in anisotropic diffusion tensor of GCR particles ($\psi$ is the angle between magnetic field lines and radial direction in the equatorial plane) and ratios $\beta=K_{\bot}/K_{\parallel}$  and $\beta_{1}=K_{d}/K_{\parallel}$  of the perpendicular $K_{\bot}$  and drift   $K_{d}$ diffusion coefficients to the parallel $K_{\parallel}$  diffusion coefficient.
A parallel diffusion coefficient used in modeling is expressed, as:
$K_{\parallel}=K_{0}K(r)K(R,\alpha)$
where $K_{0}=1.9\cdot10^{19}cm^{2}/s$, $K(r)=1+0.5r/r_{0}$;  $K(R,\alpha)=R^{\alpha}$  contributes to the changes of the parallel diffusion coefficient $K_{\parallel}$  due to dependence on the GCR particles rigidity R.
We assume that $\alpha=0.7$ in the minimum and  $\alpha=1.2$ in the maximum of solar activity.  Drift effect due to gradient and curvature of the regular IMF is implemented in the model by means of the ratio of the drift  $K_{d}$ diffusion coefficients to  the parallel $K_{\parallel}$  diffusion coefficient $\beta_{1}=K_{d}/K_{\parallel}$.
In this model, we
consider that a drift effect during the maximum solar activity is scaled down by 30\% (almost diffusion dominated case) with respect to the minimum of solar activity (drift dominated case).
The equation \ref{ParkerEq} was transformed to the algebraic system of equations using the implicit finite difference scheme, and then solved by the Gauss-Seidel iteration method. The solutions for each layer of rigidity $R$ ($R=100, 90, 80,...,10$ GV) for the stationary case are considered as an initial conditions for the nonstationary case for the given rigidity $R$ and at time $t=0$. The equation \ref{ParkerEq} in spherical coordinate system for dimensionless variables is derived in detail in \cite{Siluszyk11}, while the details of its numerical solution of the 3D nonstationary equation are given in \cite{Wawrzynczak10b}.\\
Results of the numerical solution of the equation \ref{ParkerEq} with included changes of the solar wind velocity $V$ and IMF strength $B$ (Fig.~\ref{fig:3} and Fig.~\ref{fig:4}) are presented in Fig.~\ref{fig:5} for the minimum of solar activity, and in Fig.~\ref{fig:6} - for the maximum of solar activity. The expected from the model variations for particles with rigidity $R=14$ and $10$ GV are compared with the changes in the GCR intensity observed in Oulu NM data.
To obtain more realistic results we implemented into our model in situ measurements of $V$ and $B$.  It is seen from Fig.~\ref{fig:5} and Fig.~\ref{fig:6} that we have not achieved a very good agreement between observed by Oulu NM data and modeling results neither for minimum nor maximum epochs. However, one can recognize that there are not significant distinctions between range of changes for experimental data and outcomes of theoretical calculations. Of course, using  some physically justified  assumptions, e.g., the changes in parallel and perpendicular diffusion coefficients versus the strengths $B$ and turbulence of the IMF, we could achieve more acceptable agreement between observed by Oulu NM data and modeling results. Nevertheless, our aim in this paper was to speculate, as  less as possible and create a realistic 2-D time-dependent model based on in situ observations of parameters defining the general processes in cosmic ray modulation. We believe that CD and CN distributions are related  to the processes in the cosmos. So, after careful study of the coupling  between level of ionization in lower earth atmosphere and GCR flux, CD and CN could be considered as the useful proxies of the space weather  conditions.

\section{Conclusions}
\begin{enumerate}
  \item The inter-annual distribution of cloudless days and nights in Abastumani Astrophysical Observatory reveals both annual and semi-annual variations for various levels of geomagnetic disturbances. For the annual cycle, the number of  cloudless days is the biggest in August, which may be expectable, since in this month the mean day-night surface temperature in the region is maximal. In geomagnetically disturbed conditions ($Ap\geq7$), having semi-annual character, the increase in a number of cloudless days at equinoxes possibly cause the shift of the maximal number of cloudless nights from August to September. This phenomenon and the fact that number of cloudless nights are the biggest in September and for less geomagnetic disturbances ($Ap<40$) shifts back to August, point to the influence of cosmic factors on cloud covering processes.
  \item One of the possible cosmic factors the GCRs flux change is considered, which reveals annual variations for cloudless nights, with the greatest decrease in June, where, for magnetically disturbed conditions, the maximal frequency of cloudless nights are observed.
  \item The GCRs flux observed by Oulu NM  shows a presence of some annual and inter-annual variations. These inter-annual variations differ from its modeling results but gives its noticeable reduction in June and August, where the relative number of CN and the total number of CD are increasing, respectively. This result should be important for improving an assumed model (implementing into transport equation in situ measurements of $V$ and $B$) for investigation of the observed properties of the inter-annual variations of the CD and CN distribution.

\end{enumerate}

\section*{Acknowledgments}
 We thank the principal investigator of Oulu neutron monitor  for the ability to use  data. M.V.  Alania,
 G.G. Didebulidze and M. Todua were supported by Georgian
 Shota Rustaveli National Science Foundation grant No. 13/09.

\end{document}